\documentclass{ws-ijmpa}

\newcommand{\be}{\begin{equation}}
\newcommand{\ee}{\end{equation}}
\newcommand{\ba}{\begin{eqnarray}}
\newcommand{\bc}{\begin{center}}
\newcommand{\ec}{\end {center}}

\begin{document}





\title{ATMOSPHERIC MUON FLUX AT PEV ENERGIES}

\author{S.~I.~SINEGOVSKY, A.~A.~KOCHANOV, T.~S.~SINEGOVSKAYA,}
\address{Irkutsk State University, Gagarin Blvd 20, Irkutsk, RU-664003 Russia \\ sinegovsky@api.isu.ru}



\author{A.~MISAKI,}
\address{Research Institute for Science and Technology, Waseda U., Tokyo, 169-8555, Japan
 \\ Innovative Research Organization, Saitama U., Saitama, Japan}

\author{N.~TAKAHASHI}
\address{Graduate School of Science and Technology, Hirosaki  U., Hirosaki 036-8561, Japan}

\maketitle


\begin{abstract}
 In the near future the energy region above few hundreds of TeV may really be accessible for measurements of the atmospheric muon spectrum with IceCube array.  Therefore one expects that muon flux uncertainties above 50 TeV, related to a poor knowledge of charm production cross sections and insufficiently examined primary spectra and composition, will be diminished. We give predictions for the very high-energy muon spectrum at sea level, obtained with the three hadronic interaction models, taking into account also the muon contribution due to decays of the charmed hadrons. 
 
\end{abstract}

\ccode{PACS numbers: 95.85.Ry, 13.85.Tp}


 
\section{Introduction}

The atmospheric muon flux as well as muon neutrino flux at high energies are inevitably dominated by the prompt component due to decays of the charmed hadrons ($D^\pm$, $D^0$, $\overline{D}{}^0$, $D_s^\pm$ $\Lambda_c^+,\ldots$), hence the prompt neutrino flux becomes the major source of the background in the search for a diffuse astrophysical neutrino flux\cite{nt200_06}\cdash\cite{amanda08}. Insufficiently explored processes of the charm production  give rise to most uncertainty in  the muon and neutrino fluxes. 
IceCube, the first to begin operating as the km3 neutrino telescope,  has the real capability\cite{IceCube09,Berghaus09} to measure the atmospheric muon spectrum at energies up to 1 PeV and to shed light on the feasible range of the cross sections for the charmed  particle production.

Besides, an ambiguity in high-energy behaviour of pion and kaon production cross sections
affects essentially  the atmospheric muon (neutrino) flux. 
Recent calculations\cite{kss09} reveal differences (up to factor $1.8$ at $10$ PeV) in the neutrino flux because of uncertain description of the hadronic proceses involving light quarks at high energies.

In this work we extend to higher energies the conventional muon flux calculations basing on the known hadronic interaction models with usage reliable data of the primary cosmic ray measurements. We present results of the conventional muon flux calculations in the energy range $10^5$--$10^8$ GeV using hadronic models  QGSJET-II\cite{qgsjet2,ostap06}, SIBYLL 2.1\cite{sibyll,engel99}, EPOS\cite{EPOS,EPOS2} as well as  the model by Kimel and  Mokhov\cite{KMN} (KM), that were tested also in recent atmospheric muon flux calculations\cite{KPSS,KSS08}.  
In order to compare the uncetainity of the conventional muon flux and prompt one we plot  the prompt muon contrubition originating from decays of the charmed hadrons produced in collisions of cosmic rays with nuclei of air (for review see e.g. Refs.~\refcite{bnsz89}--\refcite{DM}).
 
\section{The method  \label{sec:method}}

The high-energy muon fluxes are calculated using the approach\cite{NS} to solve the atmospheric hadron cascade equations taking into account non-scaling behavior of inclusive particle production cross-sections, rise of total inelastic hadron-nuclei cross-sections, and the non-power law primary spectrum (see also Ref.~\refcite{KSS08}). 

To obtain the differential energy spectra of protons $p(E,h)$ and neutrons $n(E,h)$ at the atmosphere depth $h$ one needs to solve the set of equations:
\begin{equation}
\frac{\partial N^{\pm}(E,h)}{\partial h}=-\frac{N^{\pm}(E,h)}{\lambda_N(E)}+
\frac{1}{\lambda_N(E)}
\int_0^1\Phi^{\pm}_{NN}(E,x)N^{\pm}(E/x,h)\frac{dx}{x^2},\label{nucleons}
\end{equation}
where $N^{\pm}(E,h)=p(E,h)\pm n(E,h)$,
\[ 
\Phi^\pm_{NN}(E,x)=\frac{E}{\sigma_{pA}^{in}(E)}\left[\frac{d\sigma_{pp}(E_0,E)}{dE}\pm
\frac{d\sigma_{pn}(E_0,E)}{dE}\right]_{E_0=E/x},
\]
 $\lambda_N(E)=1/\left[N_0\sigma_{pA}^{in}(E)\right]$  is the  nucleon interaction length in the atmosphere, $x=E/E_0$ is the fraction of the primary nucleon energy $E_0$ carried away by the secondary nucleon,
$d\sigma_{ab}/dE$ is the cross sections for inclusive reaction $a+A\rightarrow b+X$. The boundary conditions for Eq.~(\ref{nucleons}) are  $N^\pm(E,0)=p_0(E)\pm n_0(E).$

Suppose that the solution of the system is
\begin{equation}\label{ans}
N^\pm(E,h)=N^\pm(E,0)\exp\left[ -\frac{h(1-Z^\pm_{NN}(E,h))}{\lambda_N(E)}\right],	
\end{equation}
where \  \  $Z^\pm_{NN} (E,h)$ are unknown functions. 
Substituting Eq.~(\ref{ans}) into Eq.~(\ref{nucleons}) we find the equation for these functions  $Z^\pm_{NN}$ ($Z$-factors):
	\begin{equation}\label{difzf}
	\frac{\partial(hZ^\pm_{NN})}{\partial h}=\int_0^1\Phi^\pm_{NN}(E,x)\eta^\pm_{NN}(E,x)
		\exp\left[-hD^\pm_{NN}(E,x,h)\right]dx,
		\end{equation}
where $ \eta^\pm_{NN}(E,x)=x^{-2}N^\pm(E/x,0)/N^\pm(E,0),$
		\begin{equation}
		 D^\pm_{NN}(x,E,h)=\frac{1-Z^\pm_{NN}(E/x,h)}{\lambda_N(E/x)}-
		\frac{1-Z^\pm_{NN}(E,h)}{\lambda_N(E)}.
		\end{equation}	
By integrating Eq.~(\ref{difzf}) we obtain the nonlinear integral equation 
 \begin{equation}\label{zfactor}
		Z^\pm_{NN}(E,h)=\frac{1}{h}\int_0^hdt\int_0^1 dx \Phi^\pm_{NN}(E,x)\eta^\pm_{NN}(E,x)
		\exp\left[-tD^\pm_{NN}(E,x,t)\right],
		\end{equation}
which can be solved by iterations. The simple choice of zero-order approximation is $Z^{\pm(0)}_{NN}(E,h)=0$, that is 
		$D^{\pm(0)}_{NN}(E,x,h)=1/\lambda_N(E/x)-1/\lambda_N(E).$
For the $n$-th step we find
		\begin{equation}
		Z^{\pm(n)}_{NN}(E,h)=\frac{1}{h}\int_0^hdt\int_0^1 dx \Phi^\pm_{NN}(E,x)\eta^{\pm}_{NN}(E,x) 
		\exp\left[-tD^{\pm(n-1)}_{NN}(E,x,t)\right],
		\end{equation}				
where		
		\begin{equation}
D^{\pm(n-1)}_{NN}(E,x,h)=\frac{1-Z^{\pm(n-1)}_{NN}(E/x,h)}
{\lambda_N(E/x)}-\frac{1-Z^{\pm(n-1)}_{NN}(E,h)}{\lambda_N(E)}.
		\end{equation}	
		
\begin{figure}[!bh]
	\centering
\hskip 0.1 cm
\includegraphics[width=0.48\textwidth, trim = 1.0cm 0.0cm 0.5cm 0.0cm]{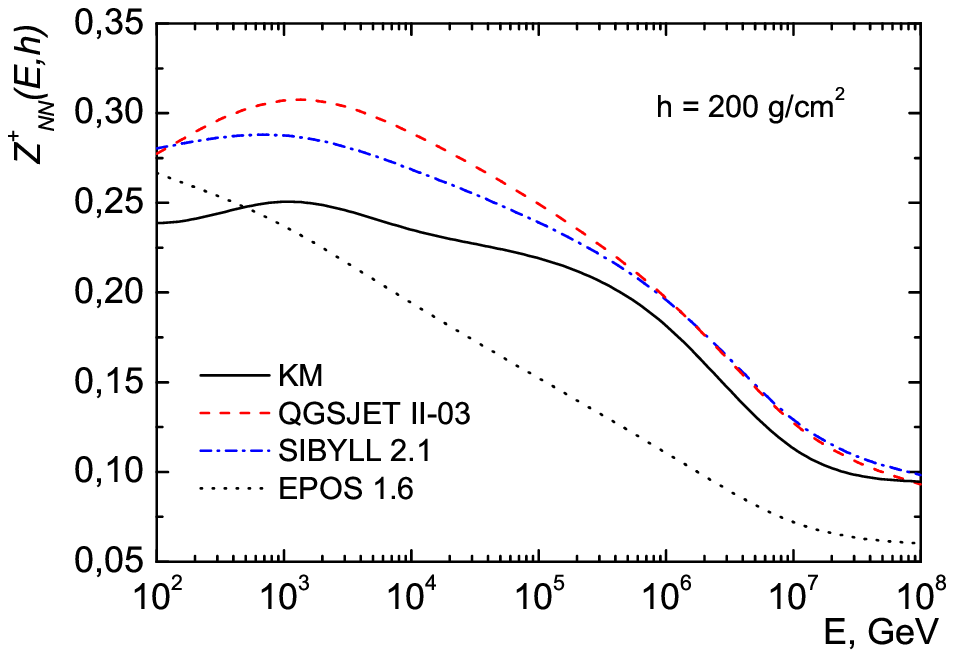}\hskip 0.30 cm
\includegraphics[width=0.48\textwidth, trim = 1.0cm 0.0cm 0.5cm 0.0cm]{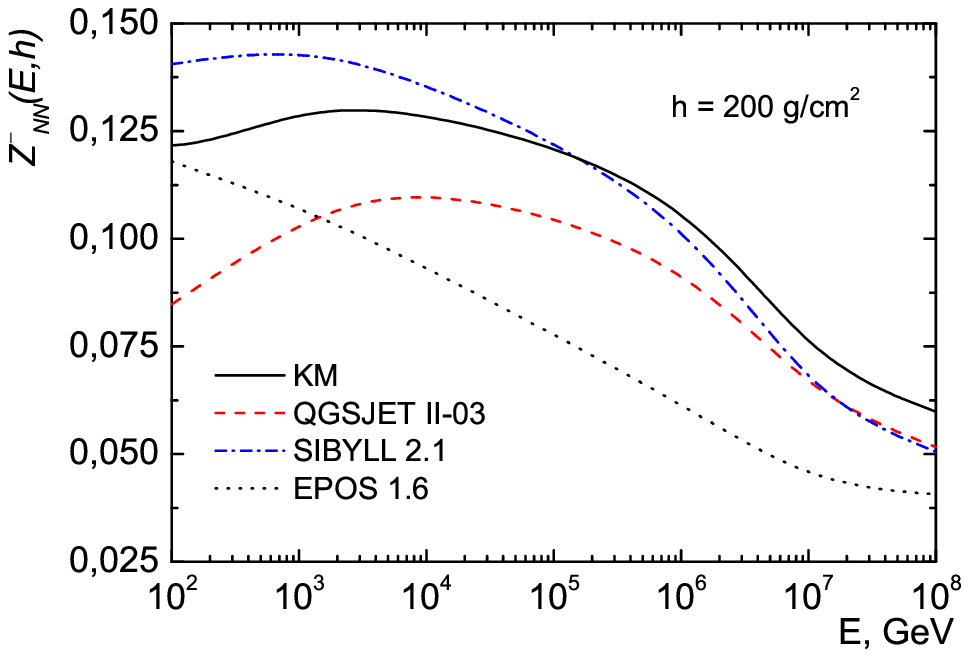}
\caption[$Z_{NN}^{\pm}$ at the depth of $200$ г$\cdot$см$^{-1}$]
{The energy dependence of the nucleon $Z_{NN}^{+}(E,h)$-function (left) and  $Z_{NN}^{-}(E,h)$  one (right) at the atmosphere depth of $200$ g$\cdot$cm$^{-2}$ calculated for the  ATIC-2 primary spectrum.}  
	\label{ZFPM}
\end{figure}

Nontrivial structure of the nucleon $Z$-factors  (Fig.~\ref{ZFPM}) results from the non-power law behavior of the ATIC-2 primary spectrum\cite{atic2}, non-scaling behavior of the particle production cross-sections and the energy dependence of inelastic hadron-nucleus cross-sections. After that, using the obtained nucleon fluxes, we are able to calculate successively the meson and the lepton fluxes (see Ref.~\refcite{KSS08} for details). 

In our calculations we rely on recent data on the primary cosmic ray (PCR) spectra and composition obtained with Advanced Thin Ionization Calorimeter in the  balloon-borne experiment ATIC-2\cite{atic2}. In order to extend the calculations to higher energies, up to $10$ PeV, we use the data of the GAMMA experiment\cite{GAMMA}. 
 The energy spectra and elemental composition, obtained in the GAMMA experiment, cover the $10^3$--$10^5$ TeV range and agree with the corresponding extrapolations of known balloon and satellite data at $E\geq10^3$ TeV. 
Alternative primary spectra, used in the calculations  for a region of very high energies, is the model by Zatsepin and Sokolskaya (ZS)\cite{ZS06,ZS07}. The ZS proton spectrum at $E\gtrsim 10^6$ GeV is compatible with KASCADE data\cite{KASCADE05,KASCADE09} and the helium  one is within the range of the KASCADE spectrum recontructed with use of QGSJET 01 and SIBYLL models. Besides  we use also the Gaisser and Honda spectra\cite{GH} (GH) to compute the muon flux in more narrow energy range $E_\mu \lesssim 10^5$ GeV.

\section{High-energy muon spectra \label{sec:an}}

Apart from evident sources of AM, $\pi_{\mu2}$ and  $K_{\mu2}$ decays, we take into consideration three-particle semileptonic decays, $K^{\pm}_{\mu3}$, $K^{0}_{\mu3}$. Also we take into account small fraction of the muon flux originated  from decay chains   $K\rightarrow\pi\rightarrow\mu$ ~  ($K^0_S\rightarrow \pi^+ + \pi^-$, $K^\pm \rightarrow \pi^\pm +\pi ^0$).

\begin{figure}[!th]
	\begin{center}
\includegraphics[width=0.85\textwidth, trim = 1.5cm 0.0cm 1.0cm 0.0cm]{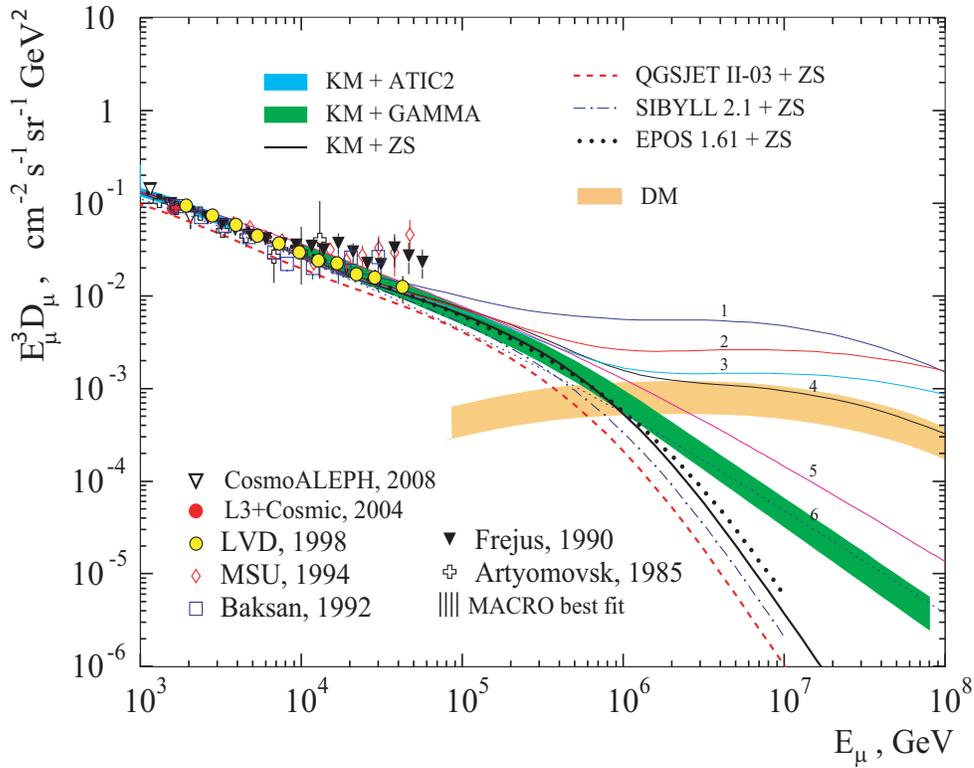}
\end{center}
	\caption{The high energy vertical muon spectra at ground level. The dashed-line curves and inclined shaded areas present this work calculations for the KM model\protect\cite{KMN} with the ATIC-2 primary spectrum\protect\cite{atic2} ($E_\mu< 10$ TeV) and GAMMA one ($E_\mu> 10$ TeV). The solid curve marks the computation for the primary spectrum model by Zatsepin and Sokolskaya\protect\cite{ZS06,ZS07}. The numbers near thin lines indicate the prompt muon flux model  ($1-4$) and the conventional muon flux ($5, 6$):  1 -- RQPM\protect\cite{bnsz89}, 2  and 3 --  two  calculation versions of PRS model\protect\cite{prs99}, 4 -- QGSM\protect\cite{bnsz89,prd98}; 5 -- KM model combined with NSU cosmic ray spectrum\protect\cite{NSU}, 6 -- the same with EKS spectrum\protect\cite{EKS}. The shaded band (DM) below the line $4$ presents the prompt neutrino flux calculation of  Ref.\protect\refcite{DM}.
}
\label{mu-2} 
   \end{figure}

The high energy spectra of the conventional and prompt muons at ground level calculated for the  vertical direction  are shown in Fig.~\ref{mu-2} together with the experimental data. 
The inclined shaded bands here indicate the conventional muon flux calculated with KM model for the case of the ATIC-2 primary spectrum  (narrow light band on the left) and the GAMMA one (dark band). The size of the bands corresponds to statistical errors in the ATIC-2 and GAMMA experiments.  Solid, dashed, dash-dotted, and dotted lines indicate the calculations  with use of  ZS spectrum and set of hadronic models, KM, SIBYLL 2.1, EPOS 1.61, and QGSJET-II 03.
The experimental data comprise the measurements of L3+Cosmic\cite{L3C}, Cosmo-ALEPH\cite{hashim07,CosmoAleph08}  as well as the data (converted to the surface) of deep underground experiments MSU\cite{MSU}, MACRO\cite{MACRO}, LVD\cite{LVD}, Frejus\cite{Frejus}, Baksan\cite{Baksan}, Artyomovsk\cite{ASD}.
Notice that the calculation results do not fit well the Frejus and MSU data even if the prompt muon component is taken into account (thin lines $1$-$4$ in Fig.~\ref{mu-2}), while the LVD data are well described.

The ZS model seems to be a reasonable bridge from TeV energy range to PeV one (solid line),  providing a junction of the different energy ranges. However, above $10^6$ GeV the muon flux is apparently affected by the primary cosmic ray ambiguity in the vicinity of  `knee'.  
To illustrate this we plot also our early predictions\cite{prd98,ts01,kss07} for the conventional muon flux made with primary cosmic ray spectra by Nikolsky, Stamenov, and Ushev (NSU)\cite{NSU} (thin line 5) as well as by Erlykin, Krutikova, and Shabelsky (EKS)\cite{EKS} (dashed line 6 over the GAMMA band).  The index  $\gamma$ of the NSU primary nucleons is $1.62$ and  $2.02$  before and beyond the  ``knee'' ($\sim 3$ PeV) correspondingly, while $\gamma=1.7$ and $2.1$ for the EKS spectrum.

The prompt muon contribution due to decays of charmed hadrons at high energies is shown here for several charm production models:  the recombination quark-parton model\cite{bnsz89} 
(RQPM, line $1$ in Fig.~\ref{mu-2}), the  two  calculation versions of the model by Pasquali, Reno and  Sarcevic\cite{prs99} (PRS, lines  $2$ and $3$), and  the quark-gluon string model\cite{bnsz89,prd98,KP85}$^-$\cite{piskunova93}(QGSM, line $4$). 
Besides we draw the shaded band (below the line $4$) calculated in Ref.~\refcite{DM}  that displays  the theoretical uncertainty for  the prompt muon flux  due to the QCD  dipole model\cite{DM} (DM). In this case we consider the prompt muon neutrino flux to be equal to the  prompt  muon one.  Also here we ignore the differences in prompt muon flux calculations related both to atmospheric hadron cascade features and the primary cosmic ray spectrum and composition.  
Both RQPM and QGSM are the nonperturbative models, whereas PRS\cite{prs99} as well as DM\cite{DM} are based on the next-to-leading-order QCD calculations. At the muon energy  above  $2$ PeV the upper edge of the DM theoretical uncertainty band is in close agreement with the QGSM prediction\cite{bnsz89,prd98}.  

\section{Muon charge ratio}

The muon charge ratio depends  on the proton to neutron ratio in primary cosmic rays as well as on the hadron production cross-sections. Thus, a comparison of the calculated $\mu^+/\mu^-$ ratio with experimental data in a wide energy range gives a possibility to study indirectly these features. At present time the muon charge ratio is measured with new facilities, that  supply with  high quality data from large number of muon events at high energies.
\begin{figure}[th!] \vspace{-0.5cm}
	\centering
\includegraphics[width=1.0\textwidth, trim = 0cm 0.0cm 0cm 0cm]{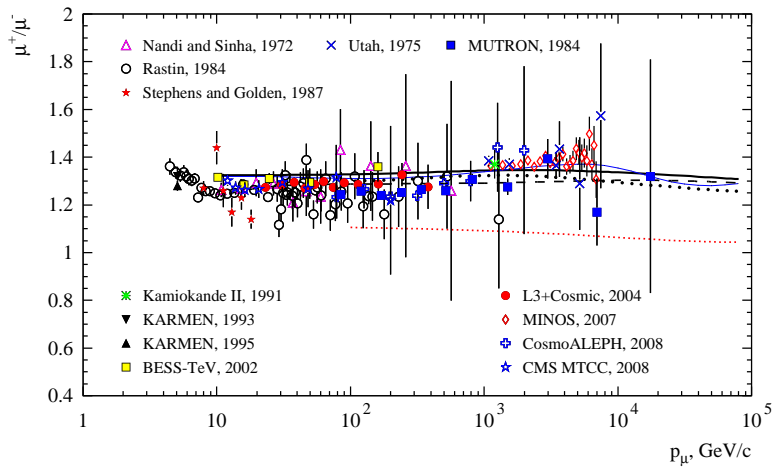} 
\caption{Muon charge ratio at ground level computed for the three hadronic interaction models and the two primary cosmic ray spectra. Solid line marks the KM + GH result for $\theta=0^{\rm o}$, dashed line shows the same at  $90^{\rm o}$. Thin line:  the KM + ZS at $0^{\rm o}$,  bold-dotted: the SIBYLL 2.1 + GH, dotted (the lower): the QGSJET-II + GH  at $0^{\rm o}$.}
  		\label{charge_ratio} 
\end{figure}

 In Fig.~\ref{charge_ratio} we present our calculations of $\mu^+/\mu^-$ ratio along with the data of  experiments\cite{L3C,CosmoAleph08,BESSTEV}\cdash\cite{CMS}. The calculations are made with the three hadronic interaction models and the  two primary cosmic ray spectra. Solid line and dashed one mark the KM + GH computation for the zenith angles $0^{\rm o}$ and $90^{\rm o}$, thin line indicates the KM + ZS result at $0^{\rm o}$. Bold dotted line indicates the  SIBYLL 2.1 + GH result, dotted shows the QGSJET-II + GH one, both for $0^{\rm o}$.
One may see that the KM and SIBYLL calculations  reproduce data well with both versions of the primary spectra,  GH and ZS. The calculated curves correspond approximately to the value $1.3$, that is in agreement with the recent data of BESS-TeV, CosmoALEPH, and  L3+Cosmic experiments up to  $1$ TeV. For higher energies the kaon source of muons becomes more intensive leading to a maximum of the muon ratio,  $\sim 1.4$,  at energy close to $10$ TeV. This value agrees with the  recent results of the MINOS far detector~\cite{MINOS}. The calculations also are in agreement 
with the spectrograph MUTRON data\cite{mutron84} at $\theta=89^{\rm o}$ including the point above $10$ TeV.  

The  QGSJET-II model (lower line) shows visible deviation from others: the predicted $\mu^+/\mu^-$ ratio is close to  $\sim 1.2$, that might be explained by the higher extent of the proton and neutron flux equalization in the atmosphere due to reactions $pA\rightarrow nX$ in the model (see Fig.~\ref{ZFPM}).

\section{Summary}

The conventional muon  spectrum in the energy region $0.1-100$ PeV  is calculated with usage of known hadronic models, QGSJET-II, SIBYLL 2.1, EPOS and KM, basing on experimental measurements of the primary cosmic ray spectrum and taking into account the ``knee''. 
We study the slope influence on tne muon spectrum in wide energy region up to the muon energy $100$ PeV,  far  beyond  the knee.  

The  calculation around the knee region (Fig.2) reveals strong dependence of the conventional muon flux on the hadronic models (involving $u,d,s$ quarks). Uncertainty in the flux due to the hadronic models reaches a half of order of magnitude, that is comparable  with the prompt muon flux uncertainty related to differences among charm production models, RQPM and QGSM or DM, for example.
One may suppose that the primary cosmic ray uncertainty beyond the knee,  as well as the  hadronic model uncertainty, would be negligible provided that real prompt muon flux at the energy around $\sim 10^6$ GeV  is not too different from the QGSM prediction (Fig.~\ref{mu-2}).           

It seems reasonable to consider that the prompt muon flux higher than it is predicted by RQPM to be excluded\cite{amanda07}. One may expect that more strong restriction of the prompt muon flux range will be extracted from the experiment in the near future.
   
\section*{Acknowledgements}  

This research was supported in part by the Russian Federation Ministry of Education and Science within the program "Development of Scientific Potential in Higher Schools" under grants 2.2.1.1/1483, 2.1.1/1539, the Federal Target Program "Scientific and educational specialists for innovative Russia", contract No. P1242.
 

\end{document}